\documentclass[twocolumn,showpacs,preprintnumbers,amsmath,amssymb,aps]{revtex4}

\pdfoutput=1
\usepackage[pdftex]{graphicx}

\usepackage{dcolumn}% Align table columns on decimal point\right] 
\usepackage{bm}% bold math

\usepackage{color}
\usepackage{amstext}
\usepackage[utf8]{inputenc}% Text encoding
\usepackage[caption=false]{subfig}% For figures side by side

\newcommand{\bra}[1] {\left<#1\right|}
\newcommand{\ket}[1] {\left|#1\right>}
\newcommand{\braket}[2] {\left<#1|#2\right>}

\newcommand{\e}[0]{\mathrm{e}}
\newcommand{\I}[0] {\mathrm{i}}

\newcommand{\mean}[1] {\left<#1\right>}

\begin{document}

%%%%%%%%%%%%%%%%%%%%%%%%%%%%%%%%%%%%%%%%%%%%%%%%%%%%%%%%%%%%%%%%%%%%%%%%%%
%                               Title                                    %
%%%%%%%%%%%%%%%%%%%%%%%%%%%%%%%%%%%%%%%%%%%%%%%%%%%%%%%%%%%%%%%%%%%%%%%%%%
\title{Simple approach for the two-terminal conductance through interacting clusters}
\author{A. A. Lopes}

\affiliation{Institute of Physics, University of Freiburg, Hermann-Herder-Straße 3, \\
79104 Freiburg, Germany.}

\author{R. G. Dias}
\affiliation{Department of Physics, I3N, \\
University of Aveiro, \\
Campus de Santiago, Portugal}

\date{\today}

%%%%%%%%%%%%%%%%%%%%%%%%%%%%%%%%%%%%%%%%%%%%%%%%%%%%%%%%%%%%%%%%%%%%%%%%%%
%                              abstract                                  %
%%%%%%%%%%%%%%%%%%%%%%%%%%%%%%%%%%%%%%%%%%%%%%%%%%%%%%%%%%%%%%%%%%%%%%%%%%
\begin{abstract}
We present a new method for the determination of the two-terminal differential  conductance through an interacting cluster, where one maps the interacting cluster   into a non-interacting cluster of $M$ independent sites (where $M$ is the number of cluster states with one  particle  more or less than the ground state of the cluster), with different onsite energy and connected to the leads with renormalized hoppings constants. The onsite energies are determined from the one-particle (one-hole) excitations of the interacting cluster and the hopping terms are given by the overlap between the interacting $N$ particle ground state and the one-particle (one-hole) excitations of the interacting cluster with $N$-1 ($N$+1) particles. The conductance is obtained from the solution of a system of $M$+2 coupled linear equations.
We apply this method to the case of the conductance of spinless fermions through an AB$_2$ ring taking into account nearest neighbors interactions. We discuss the effects of interactions on the zero frequency dipped conductance peak characteristic of the non-interacting AB$_2$ ring as well as the consequences of a particle number jump that occurs as the gate potential is varied.
\end{abstract}

\maketitle

%%%%%%%%%%%%%%%%%%%%%%%%%%%%%%%%%%%%%%%%%%%%%%%%%%%%%%%%%%%%%%%%%%%%%%%%%%
%                             Introduction                               %
%%%%%%%%%%%%%%%%%%%%%%%%%%%%%%%%%%%%%%%%%%%%%%%%%%%%%%%%%%%%%%%%%%%%%%%%%%

\section{Introduction}

The signatures of electronic interactions in the conductance through  nanosystems  has drawn a great deal of attention in the past few years. Many conductance studies have addressed the interaction effects using the method of  non-equilibrium Keldysh Green functions.\cite{keldysh_diagram_1965,meir_landauer_1992} In this approach, one assumes non-interacting leads   and  the  non-equilibrium current is obtained  as a function of the exact propagators of the interacting cluster (including the contribution of the leads). 
When the coupling between the non-interacting leads and the interacting cluster is small, a simpler approach by Jagla and Balseiro \cite{jagla_electron-electron_1993} which requires only the determination of the Green's functions of the decoupled interacting cluster can be used.  This approach maps the scattering through the interacting cluster  into a Landauer-type formula which relates the electrical resistance to the
one-particle scattering properties of an effective impurity. 
Note however that this approach does not capture the correlation effects that lead to the Kondo phenomena.

In this paper, a new method (that reproduces the results in Jagla's and Balseiro's approach) for the  two-terminal conductance through interacting clusters is proposed. This method relies in the approximation that despite interacting with other particles in the scattering region, the incident particles  remain independent   in the leads. This assumption allows us to work in a reduced Hilbert space and the conductance is obtained from the solution of a system of $M$+2 coupled linear equations (avoiding the need to calculate Green's functions), where $M$ is the number of cluster states with one particle more or less relatively to the ground state of the cluster.  

As an application of this method, 
we address the conductance of spinless fermions through an interacting AB$_2$ ring.
The AB$_2$ chain, also known as \emph{diamond chain}, \cite{macedo_magnetism_1995,montenegro-filho_doped_2006} is a prime example of the family of itinerant geometrically frustrated systems \cite{vidal_aharonov-bohm_1998,tamura_flat-band_2002,tasaki_nagaokas_1998,mielke_ferromagnetism_1999,derzhko_structural_2005,tanaka_metallic_2007,duan_theoretical_2001,richter_exact_2004,wu_flat_2007,wu_p_xy-orbital_2008}, which include
among others, the well known Lieb's and Kagome's lattice \cite{derzhko_low-temperature_2010}. These systems exhibit the interesting feature that they show one or more flat bands in their energy dispersion relation. As for the AB$_2$ chain, this interesting system has been heavily studied lately and several results including exact solutions for the ground state of an Hubbard AB$_2$ chain \cite{gulacsi_exact_2007}, the exact solution in the case of interacting spinless fermions \cite{lopes_interacting_2011} and the exact solution in the case of the distorted Ising-Hubbard AB$_2$ chain \cite{nalbandyan_magnetic_2014} have been obtained. There have also been exhaustive studies in the case of an extended Hubbard AB$_2$ chain \cite{rojas_geometrical_2012} as well as studies concerning the effect of disorder on the AB$_2$ chain \cite{leykam_flat_2013} and several experimental studies in materials exhibiting the same geometry \cite{kikuchi_experimental_2005, rule_nature_2008, honecker_dynamic_2011} of which azurite is a prime example.
The conductance through the non-interacting AB$_2$  chain has been recently studied by us  \cite{lopes_conductance_2014}  and   unusual features were found. In particular, a zero frequency dipped conductance peak is present due to the existence of localized states with nodes in the probability density.
In this paper, we study the evolution with increasing nearest neighbor interaction  of the conductance profiles of the AB$_2$ ring and show that 
the dipped peak  persists for small values of the interaction, but slightly shifted from zero frequency. Also, as the gate potential is varied,  one observes a  particle number jump in the cluster that filters  some conductance peaks. This particle number jump  can be associated with the flat band of the AB$_2$ ring when interactions are absent.

This paper is organized in the following way. In section II, we discuss the approximations that underlie our mapping of the interacting system into a one-particle non-interacting system. The details of the corresponding   analytical calculation can be found in appendix \ref{app:DFC}.
In section III, we apply the method to the determination of the conductance through  an interacting AB$_2$ ring.
In section IV, we conclude.
The exact solution of the  AB$_2$ Hamiltonian in the non-interacting limit as well as in the strong coupling limit is reviewed in appendix \ref{app:ESAB2C}.

%%%%%%%%%%%%%%%%%%%%%%%%%%%%%%%%%%%%%%%%%%%%%%%%%%%%%%%%%%%%%%%%%%%%%%%%%%
%                       Conductance in the diamond star                  %
%%%%%%%%%%%%%%%%%%%%%%%%%%%%%%%%%%%%%%%%%%%%%%%%%%%%%%%%%%%%%%%%%%%%%%%%%%
\section{Conductance through interacting clusters}

\label{sec:CTIC}
\begin{figure}[tp]
    \includegraphics[width=8cm]{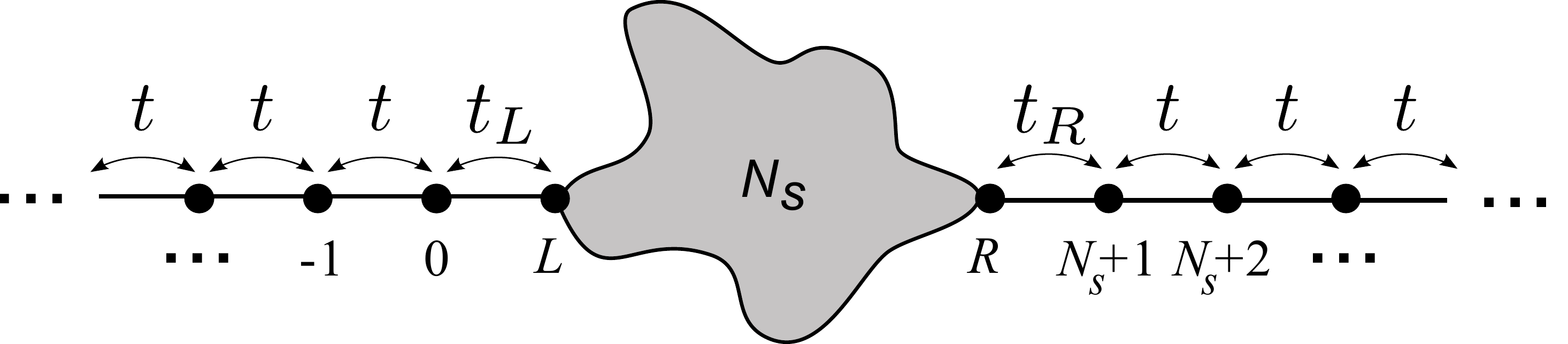}
    \caption{An interacting cluster with $N_s$ sites connected to two semi-infinite one-dimensional tight-binding leads at sites $L$ and $R$ ($L,R \in \lbrace 1, \ldots N_s \rbrace$).}
    \label{fig:conductanceSys}
\end{figure}

We consider a cluster with sites $L$ and $R$ connected to left and right leads respectively. We consider these leads to be described by a one-dimensional tight binding model and to be weakly coupled to our cluster as depicted in Fig. \ref{fig:conductanceSys}. The Hamiltonian of the full system is given by
\begin{equation}
    H = H_{C} + H_{\text{leads}} + H_{LR},
\end{equation}
where $H_{C}$ is the Hamiltonian of the isolated cluster (an AB$_2$ ring, in the case of Fig.~\ref{fig:DiamondStar4SidedLeads}),  $H_{\text{leads}}$ is the Hamiltonian of the isolated leads, assumed to be semi-infinite,
\begin{equation}
    H_{\text{leads}} = -t \sum_{j=1}^\infty (a_{j}^\dagger a_{j+1} + a_{-j-1}^\dagger a_{-j}) + \text{H.c.},
\end{equation}
and the hybridization term is
\begin{equation}
    H_{LR} = - t_L a^\dagger_{-1} X_{L} -t_R  a^\dagger_{1} X_{R} + \text{H.c.},
\end{equation}
where $t_L$ and $t_R$ are  the hopping amplitudes coupling the leads and the cluster and $X_{L}^\dagger$ and $X_{R}^\dagger$ create an electron on site L and R, respectively, of the cluster.
For simplicity, we have  assumed spinless fermions,  but our approach can be generalized to the  spinful case as well as to a multi-terminal configuration.

%%%%%%%%%%%%%%%%%%%%% 
%      figure       % 
%%%%%%%%%%%%%%%%%%%%%
%%%%%%%%%%%%%%%%%%%%% 
%      end          % 
%%%%%%%%%%%%%%%%%%%%%

Let us discuss first the non-interacting case. In this case, each energy value in the band continuum of the semi-infinite leads is twice degenerate. If $t_L = t_R =0$, and assuming  incoming particles only from one lead, this corresponds to  states such that  an incoming particle is totally reflected in the  left lead or in the right lead. When  $t_L, t_R \neq 0$, the presence of the cluster generates a certain  mix of these states (again  assuming  incoming particles only from one lead). In fact, apart from the cluster, this particle wave function with energy in the band continuum of the semi-infinite leads remains a combination of incident, reflected and transmitted plane waves.
Eigenstates with energies outside the band continuum are localized states in the cluster.
Note that this is valid for any value of $t_L$ and $t_R$ if the cluster is finite and the leads are semi-infinite, and this can be understood noting that any finite localized term in the Hamiltonian is an infinitesimal perturbation in the plane wave basis, if the leads are infinite.

%%%%%%%%%%%%%%%%%%%%% 
%      figure       % 
%%%%%%%%%%%%%%%%%%%%% 
\begin{figure}[t]
    \includegraphics[width=.4 \textwidth]{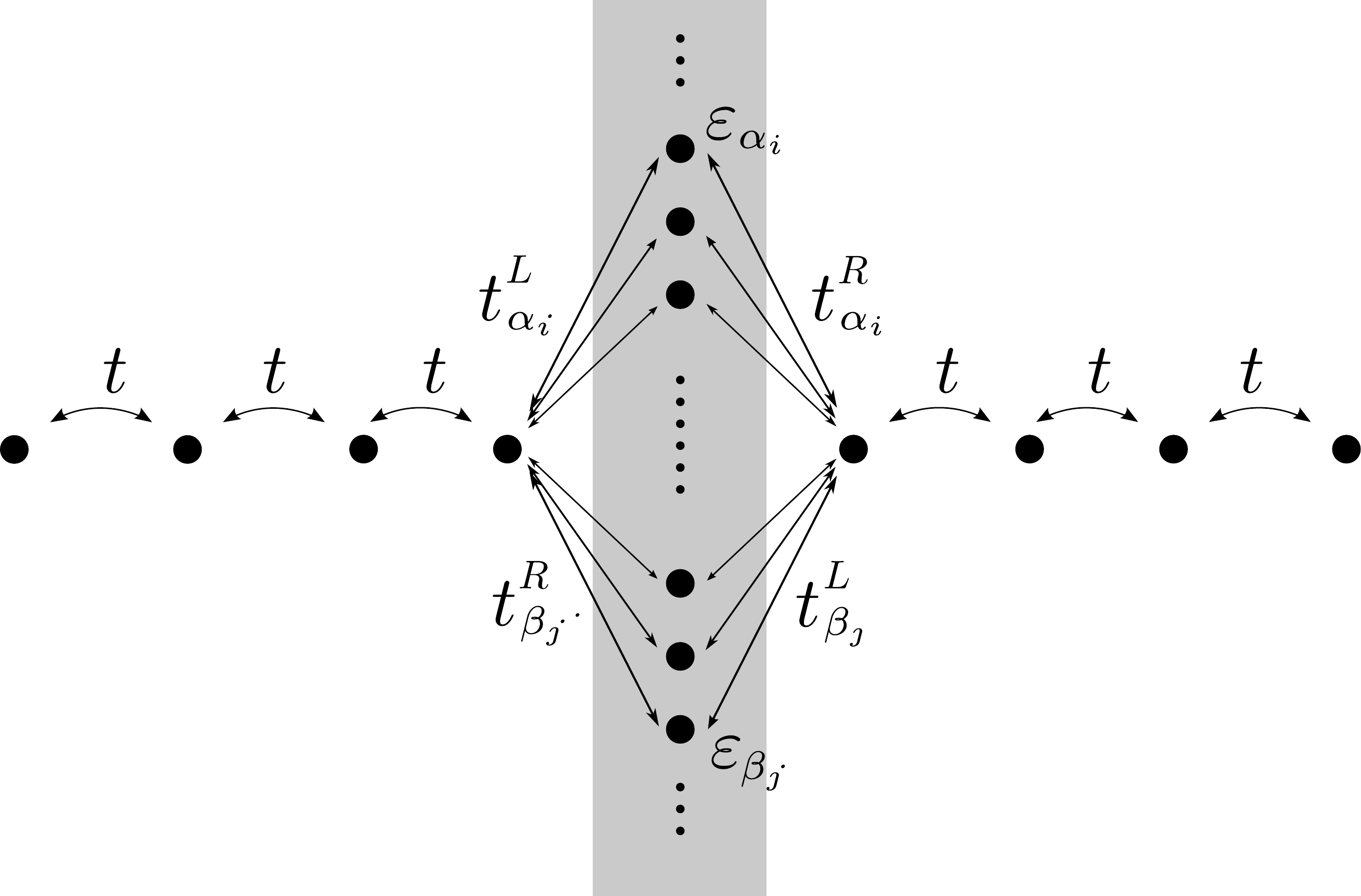}
    \caption{Non-interacting cluster obtained in our method.
    The onsite energies $\varepsilon_{\alpha_i}$ ($\varepsilon_{\beta_j}$) are determined from the one-particle (one-hole) excitations of the interacting cluster and the hopping terms $t^{L/R}_{\alpha_i}$ ($t^{L/R}_{\beta_j}$) are given by the overlap between the interacting $N$ particle ground state and the one-particle (one-hole) excitations of the interacting cluster with $N+1$ ($N-1$) particles. See appendix A for the exact expressions of these parameters.}
    \label{fig:effectcluster}
\end{figure}
%%%%%%%%%%%%%%%%%%%%% 
%      end          % 
%%%%%%%%%%%%%%%%%%%%%

When interactions are present in the cluster, we assume  that  the same occurs, that is, that the presence of the interacting cluster will merely mix the non-interacting one-particle states of the left and right leads with same energy  and  that  particles in the leads remain independent despite interacting with other particles in the scattering region  (this is also implicit in Balseiro's approach) and therefore we can study the  transmission of a single incoming particle following a Landauer-like procedure.
Furthermore, the cluster  state remains the same as in the decoupled situation  when the incoming particle is far 
from the cluster since the number of particles in the cluster is fixed by the chemical potential in the leads (we consider an infinitesimal chemical potential difference between the leads).
Therefore the incoming particle will arrive at the contact site, with the cluster at its interacting ground state, and in order for transmission to occur the system must be able to have a transition into the state where a particle is at the right contact site and the cluster is again in the ground state.
There are two possible paths for such transition, one path involving an intermediate  state with $N+1$ particles in the cluster and zero particles in the leads and another path involving an intermediate  state $N-1$ particles in the cluster and one particle at site 0 and one particle at site $N_s+1$, where $N_s$ is the number of sites of the cluster (see Fig.~\ref{fig:conductanceSys}). 
This reasoning allows us to work in a reduced Hilbert space and the conductance is obtained from the solution of a system of $M+2$ coupled linear equations, where $M$ is the number of cluster states with one extra particle or one less relatively to the ground state of the cluster. 
So, we reduce the determination of the conductance through an interacting cluster to a one-particle transmission problem through a non-interacting cluster but with hopping constants and local energies determined taking into account the  interactions in the  cluster. The analytical details  of this approach can be found in appendix A.
The relation between the transmission probability and the conductance is given by  the usual Landauer formula \cite{landauer_electrical_1970}.

In Fig.~\ref{fig:effectcluster}, we show the  non-interacting cluster obtained in our method, with  $M$  decoupled sites, with different onsite energy 
($\varepsilon_{\alpha_i}$ for sites $\alpha_i$ corresponding to states with $N+1$ particles and $\varepsilon_{\beta_j}$ for sites $\beta_j$ corresponding  to states  with $N-1$ particles) 
and connected to the leads with renormalized hoppings constants ($t^{L/R}_{\alpha_i}$ for sites $\alpha_i$ and $t^{L/R}_{\beta_j}$ for sites  $\beta_j$).
The onsite energies are determined from the one-particle (one-hole) excitations of the interacting cluster and the hopping terms are given by the overlap between the interacting $N$ particle ground state and the one-particle (one-hole) excitations of the interacting cluster with $N-1$ ($N+1$) particles. Note that these onsite energies and renomalized hoppings do not depend on the energy of the incident particles (as in Jagla's and Balseiro's approach \cite{jagla_electron-electron_1993}), but only on the number of particles in the ground state of the cluster. Note also the exchange of the indices $L$ and $R$ in the case of hoppings to $\beta$ sites, reflecting the fact that transmission of an incoming particle through these states involves first a particle hop from the cluster to the right lead and second, the hopping of the incoming particle from the left lead to the cluster.

%Differential conductance

One can ask what happens when the ground state of the interacting cluster  is degenerate.  Some authors have avoided this problem assuming the existence of a small perturbation  that lifts the ground state degeneracy \cite{rincon_features_2009}. Depending on this perturbation,   some conductance peaks may however disappear from the conductance profile. In this paper, we use a different approach. Since we are addressing the differential conductance, the chemical potential of the cluster is well defined, and the distribution of cluster states at finite temperature will be given by the density operator for the grand canonical ensemble,
\begin{equation}
    \rho = \dfrac{\e^{-\beta (H-\mu)}}{\mathrm{tr}\, \e^{-\beta (H-\mu)} },
\end{equation}
and the average current will, consequently, be given by
\begin{equation}
    \mean{I} := \mathrm{tr} \rho I.
\end{equation}
At zero temperature, the sum will be over only ground states
\begin{equation}
    \mean{I} = \dfrac{\sum_{i} \bra{E_\text{G.S.}^{(i)} } I \ket{E_\text{G.S.}^{(i)}} }{\sum_{i} \braket{ E_\text{G.S.}^{(i)} } {E_\text{G.S.}^{(i)} } },
\end{equation}
such that the average conductance is simply the arithmetic mean of the conductances for each ground state.
This way, all conductance peaks associated with the several ground states will be present in the conductance profile. However, note these peaks may have the  height reduced due to the averaging, if they are associated with only one ground state. Even if every ground state generates a certain conductance peak, the peak may have a different width for each cluster ground state (due to different values of the  effective hopping to these states, see Fig.~\ref{fig:effectcluster}) and the averaging will generate a peculiar non-Lorentzian peak with a sharper maximum.

%%%%%%%%%%%%%%%%%%%%%%%%%%%%%%%%%%%%%%%%%%%%%%%%%%%%%%%%%%%%%%%%%%%%%%%%%%
%                       Conductance in the diamond star                  %
%%%%%%%%%%%%%%%%%%%%%%%%%%%%%%%%%%%%%%%%%%%%%%%%%%%%%%%%%%%%%%%%%%%%%%%%%%
\section{Conductance in the AB$_2$ chain}

\begin{figure}[t]
    \includegraphics[width=8cm]{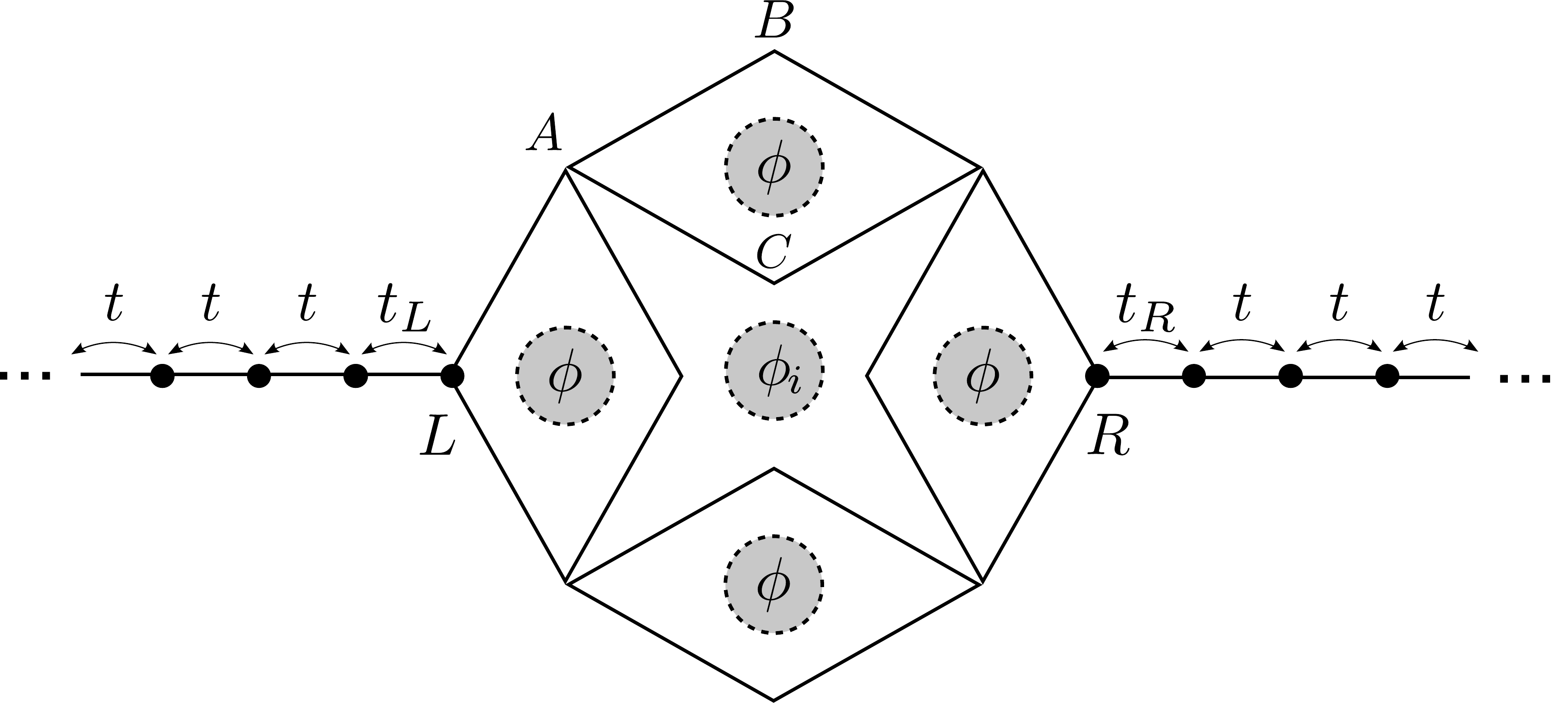}
    \caption{The AB$_2$ chain is connected at sites $L$ and $R$, to one-dimensional semi-infinite tight binding leads via  hopping amplitudes $t_L$ and $t_R$. Each plaquette is threaded by a magnetic flux $\phi$ while the inside ring is threaded by a magnetic flux $\phi_i$.}
    \label{fig:DiamondStar4SidedLeads}
\end{figure}

In this section, we apply the method described in section II (and detailed in appendix A) to the study of the conductance through an AB$_2$ ring (see Fig.~\ref{fig:DiamondStar4SidedLeads}).
Recently, we have shown that due to the existence of one-particle localized states in this ring (consequence of the  geometrical frustration), the conductance through the non-interacting  AB$_2$ ring displays a peculiar dipped conductance peak at zero frequency.
Here, we discuss the effect of the nearest-neighbor interactions on the presence of this peak as well as effects due to 
a particle number jump [due to the existence of a flat band when no interactions are present (see appendix B), but that persists when interaction is taken into account] which occurs as a gate voltage $V_g$ is varied.

%%%%%%%%%%%%%%%%%%%%% 
%      figure       % 
%%%%%%%%%%%%%%%%%%%%% 
%%%%%%%%%%%%%%%%%%%%% 
%      end          % 
%%%%%%%%%%%%%%%%%%%%%

%%%%%%%%%%%%%%%%%%%%% 
%      figure       % 
%%%%%%%%%%%%%%%%%%%%% 
\begin{figure*}[t]
    \subfloat [$\phi=0$] 
    {\label{fig:densityPlotLegend}\includegraphics[width=.45 \textwidth]{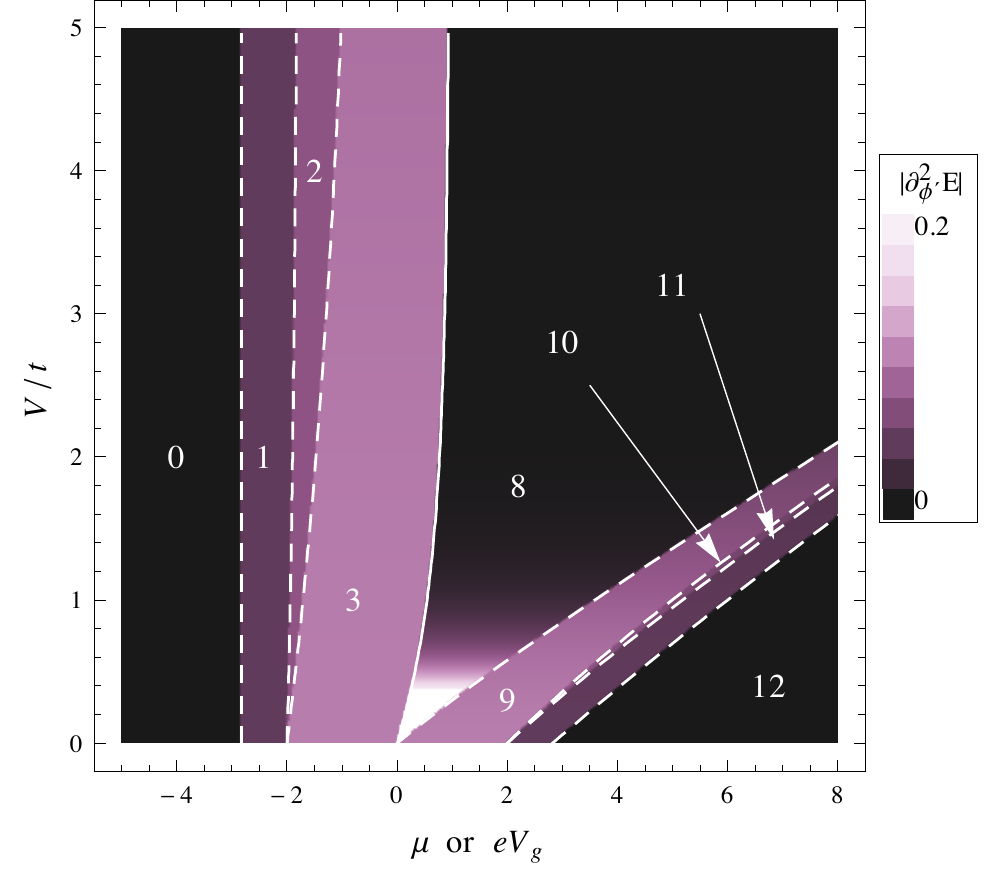}}
    \subfloat [$\phi=\pi/2$] 
    {\label{fig:densityPlotLegendwithflux}\includegraphics[width=.45 \textwidth]{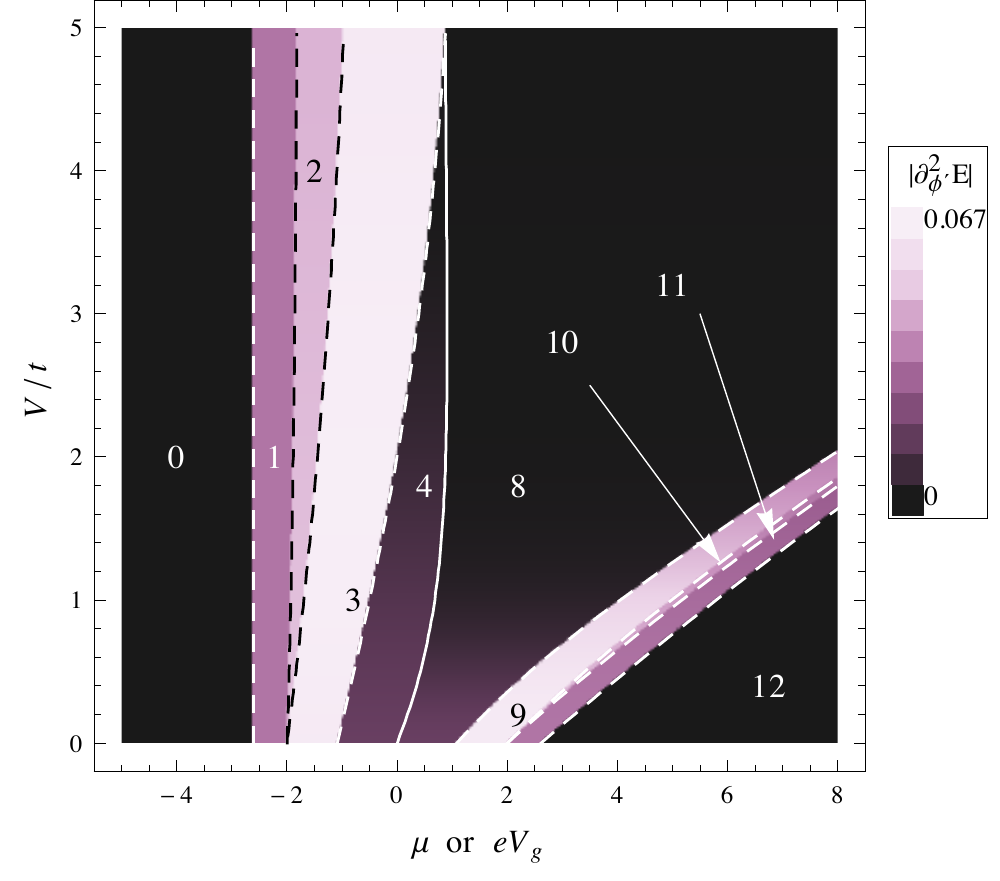}}   

    \caption{Phase diagram of the AB$_2$ ring connected to a particle reservoir as a function of  chemical potential and  interaction for  $N_c = 4$ and (a) zero flux or (b) flux per plaquette $\phi=\pi/2$. The white and black  lines delimit areas of different number of particles on the ground state while the density plot gives us the magnitude of the second derivative of the ground state energy with respect to the flux $\phi'$ at $\phi' = 0$ (proportional to the the charge stiffness). The numbers indicate the number of particles in the ground state. The results were obtained using forward differences and a step size  $d\phi = \pi \times 10^{-4}$.}
    \label{fig:phaseDiag}
\end{figure*}
%%%%%%%%%%%%%%%%%%%%% 
%      end          % 
%%%%%%%%%%%%%%%%%%%%%
In Fig.~\ref{fig:DiamondStar4SidedLeads}, an AB$_2$ ring is shown with a magnetic flux $\phi$ threading each plaquette and a magnetic flux $\phi_i$ threading the inner ring. 
 The  inner sites in the AB$_2$ ring of Fig.~\ref{fig:DiamondStar4SidedLeads} are denoted as C sites and the  outer sites as B sites. Spinal sites are denoted as A sites. The number of unit cells (or plaquettes) is denoted $N_c$.
Therefore, the magnetic flux enclosed by the outer ring is
 $\phi_o = \phi_i + 4N_c \phi/4$ and   we  introduce an auxiliary flux $\phi^\prime$ such that $\phi_o = \phi^\prime + 2N_c \phi/4$, $\phi_i = \phi^\prime - 2N_c \phi/4$.

The Hamiltonian for an AB$_2$ ring with $N_c$ unit cells is
\begin{equation}
    H_C = H_0 + V \sum_j \left(n^A_j + n^A_{j+1}\right)\left( n^B_j + n^C_j \right),
    \label{AB2}
\end{equation}
where $V$ is the value of the interaction and 
\begin{equation}
\begin{split}
    H_0 = -t \sum_{j=1}^{N_c} &   \left[ \e^{\I \phi_o/2N_c} ( A_j^\dagger B_j  + B_j^\dagger A_{j+1}) \right. \\
    & +  \left. \e^{-\I \phi_i/2N_c} ( C_j^\dagger A_j + A_{j+1}^\dagger C_j ) \right] + \text{H.c.},
\end{split}
\end{equation}
where $A_j^\dagger$ creates a particle at the spinal site of the unit cell $j$ of the AB$_2$ ring, and $B_j^\dagger$  and $C_j^\dagger$ creates a particle at the edge sites (see Fig.~\ref{fig:DiamondStar4SidedLeads}).
Here we have chosen a gauge such that the Peierls phases are equally distributed in the inner and outer ring of the  AB$_2$ ring.
In order to  introduce a gate voltage in our cluster one needs to modify the cluster Hamiltonian, $H_C \rightarrow H_C - eV_gN$.

%%%%%%%%%%%%%%%%%%%%% 
%      figure       % 
%%%%%%%%%%%%%%%%%%%%% 
\begin{figure}[bt]
%\hspace{1cm}
    \subfloat [$V/t=0.1$] 
    {\label{fig:GexpresultsV01}\includegraphics[width=.45 \textwidth]{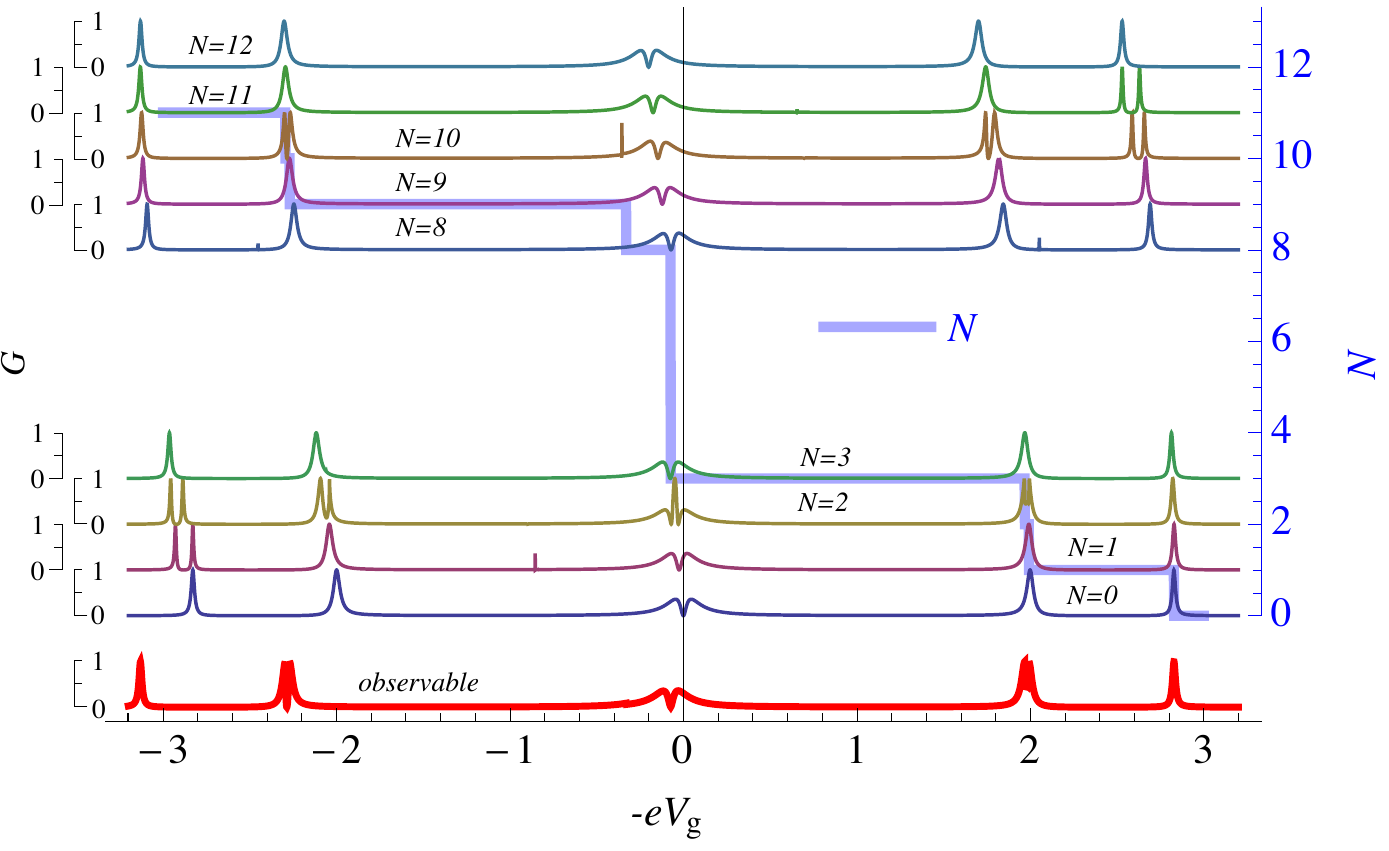}}\\
    \subfloat [$V/t=1$] 
    {\label{fig:GexpresultsV1}\includegraphics[width=.45 \textwidth]{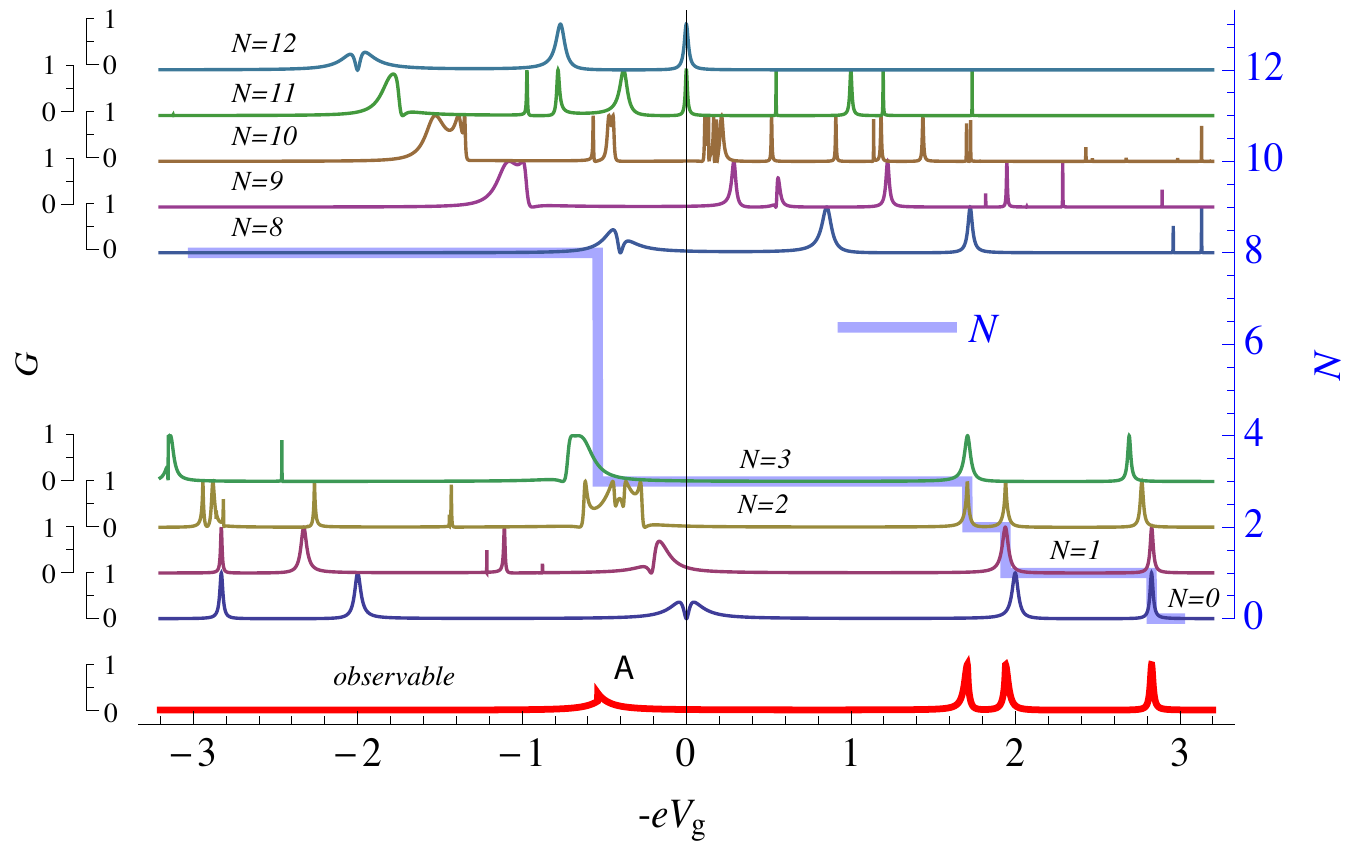}}\\
    \subfloat [$V/t=100$] 
    {\label{fig:GexpresultsV100}\includegraphics[width=.45 \textwidth]{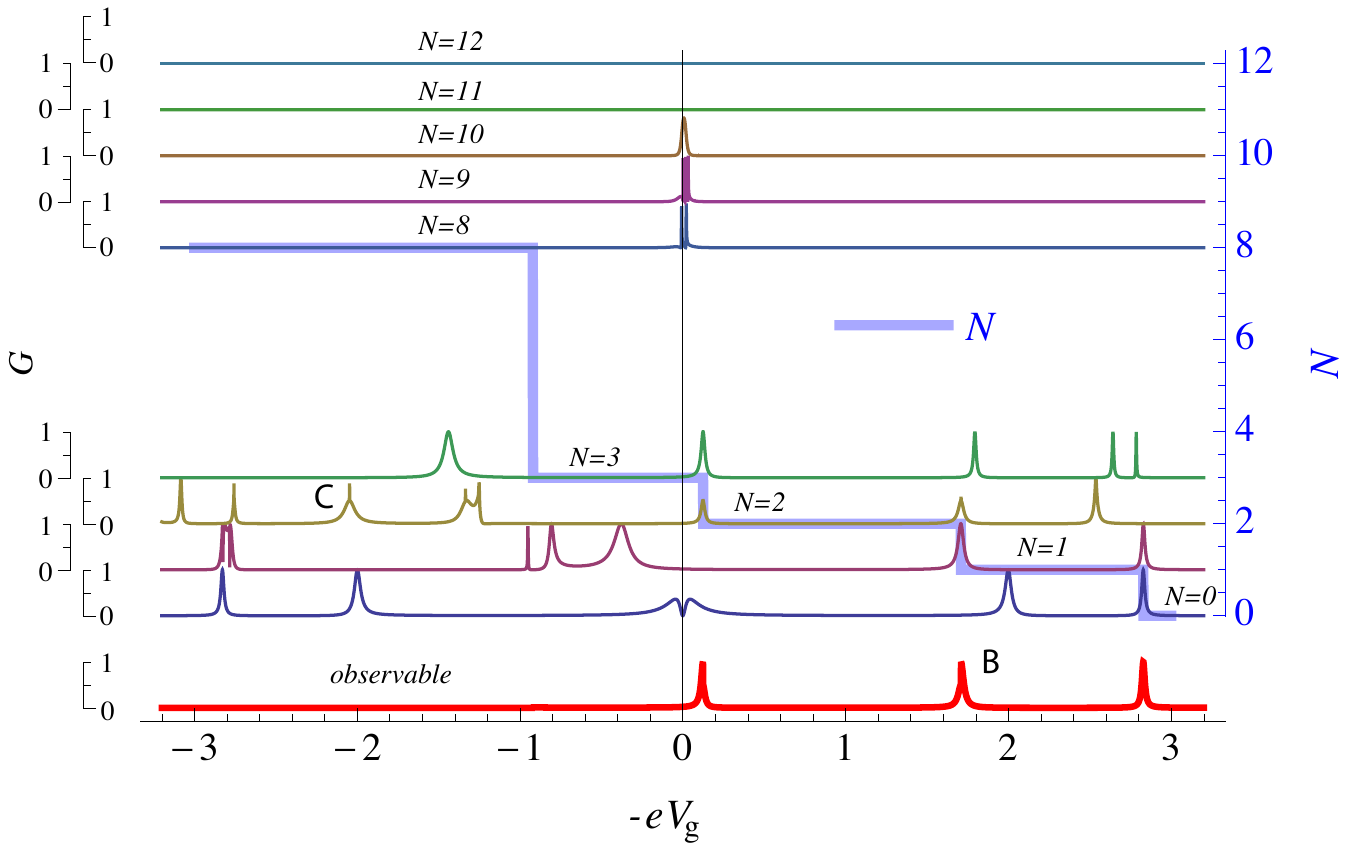}}
    \caption{Conductance  as function of -$e V_g$  for several values of the interaction constant, (a) $V/t=0.1$, (b) $V/t=1$ and (c) $V/t=100$, with contacts at sites B and C of the same unit cell and for zero flux. The conductance profiles for fixed particle number in the cluster are shown as well as the observable conductance (bottom red line) obtained taking into account the particle number transitions (blue thick line, right axis).}
    \label{fig:Gzeroflux}
\end{figure}
%%%%%%%%%%%%%%%%%%%%% 
%      end          % 
%%%%%%%%%%%%%%%%%%%%%

In the case of a transport experiment, the AB$_2$ ring is connected to particle reservoirs and one needs to know the ground state of the AB$_2$ ring  as function of  chemical potential.
The phase diagram of an AB$_2$ ring with four unit cells (as in Fig.~\ref{fig:DiamondStar4SidedLeads}) as a function of  chemical potential and  interaction  is displayed in Fig.~\ref{fig:densityPlotLegend} for zero flux and in Fig.~\ref{fig:densityPlotLegendwithflux} for flux per plaquette equal to $\pi/2$.

For zero flux, the  phase diagram Fig.~\ref{fig:densityPlotLegend} shows a particle number jump to a high density state, from 3 particles to 8 particles (or better, from $N_c$-1 to 2$N_c$ particles).
It is interesting to note that the cluster is never at half-filling or close to it. As that is the situation corresponding to the largest Hilbert subspace, the fact that one can neglect these states means that our computational effort is reduced. This particle jump can be associated with  the existence of zero energy localized states (due to the geometric frustration of the AB$_2$ chain) when $V/t$ is zero (see appendix B), since  a small chemical potential shift around $\mu =0$ implies the immediate fill of the respective  flat band (which separates two bands of itinerant states).
Curiously this particle jump survives for any value of  $V/t$. This is not obvious since we do not expect  the states of the flat band to remain localized when $V$ is finite, if the lower itinerant band is full. 

In Fig.~\ref{fig:phaseDiag} we also show  as a density plot  the charge stiffness at zero temperature (or better, the curvature of the ground state energy) as a function of  chemical potential and interaction.
As  first stated by Kohn, \cite{kohn_theory_1964,zotos_evidence_1996}  in 1D systems  the charge stiffness at zero temperature can be obtained from the ground state energy dependence on the  magnetic flux and this allows one 
 to distinguish an ideal insulating ground state from a ideal metallic one. In the case of the AB$_2$ ring, the change of boundary conditions is related to the variation of flux $\phi'$ and the charge stiffness is given  by
\begin{equation}
    D_c =N_c \left. \dfrac{\partial^2 E(\mu,V)}{\partial \phi^{\prime 2}} \right\vert_{\phi'=0},
\end{equation}
where $E(\mu,V)$ is the many-body ground state energy of the AB$_2$ ring with chemical potential $\mu$ and interaction $V$.The flux $\phi^\prime$ is analogous  to the flux threading a normal quantum ring. At zero temperature, one expects $D_c = 0$ for an insulating state and $D > 0$ for a metallic one.
Note that for fixed $\phi$, a derivative in order to $\phi'$ is the same as a derivative in order to $\phi_i$ or $\phi_o$.

In Fig.~\ref{fig:densityPlotLegend}, one can  see that while an increase in $V$ has a small impact on the charge stiffness  for most values of $\mu$, for 8 particles, a small increase in the interaction immediately turns the system into an insulator.
This reflects the fact that for 8 particles in the cluster and strong interaction, the ground state corresponds to a Wigner crystal configuration,  with all B and C sites occupied (all particles are localized due to the interaction).

When the magnetic flux threading the AB$_2$ plaquettes is $\pi/2$, the  phase diagram (Fig.~\ref{fig:densityPlotLegendwithflux}) displays an additional region corresponding to 4 particles in the AB$_2$ ring, and the  particle number jump occurs between from 4 to 8 particles (or better, from $Nc$ to $2Nc$ particles). This region becomes narrower as the interaction grows and disappears for $V/5 \sim 1$.
The existence of this region is justified by the fact that 
a finite flux $\phi$ through each plaquette induces a gap between the itinerant bands and the localized band when $V/t=0$. Therefore one has  a finite chemical potential interval (corresponding to this gap), where the lower itinerant band is completely full. This behavior remains for finite values of the interaction as long  as the interaction is small compared with the gap.
This does not happen for $\phi=0$ since in that case the energy of highest level of the lower itinerant band is zero and coincides with the flat band.
We note that the qualitative behavior for other number of cells $N_c$ is identical to the one studied above. The only difference is related to the parity of $N_c$. For $\phi = 0$ and an even number of cells, the particle jump occurs from $N_c - 1$ to $2N_c$ as for 4 cells. However for $N_c$ odd, since no single particle state in the lower band coincides with those of the flat band, there is an effective gap between these two bands, and one observes that the particle jump occurs from $N_c$ to $2N_c$ even for $\phi=0$, for small values of the interaction $V$.

\begin{figure}[bt]
%\hspace{1cm}
    \subfloat [$V/t=0.1$] 
    {\label{fig:GexpresultsV01_B1C1_fluxPiover2}\includegraphics[width=.45 \textwidth]{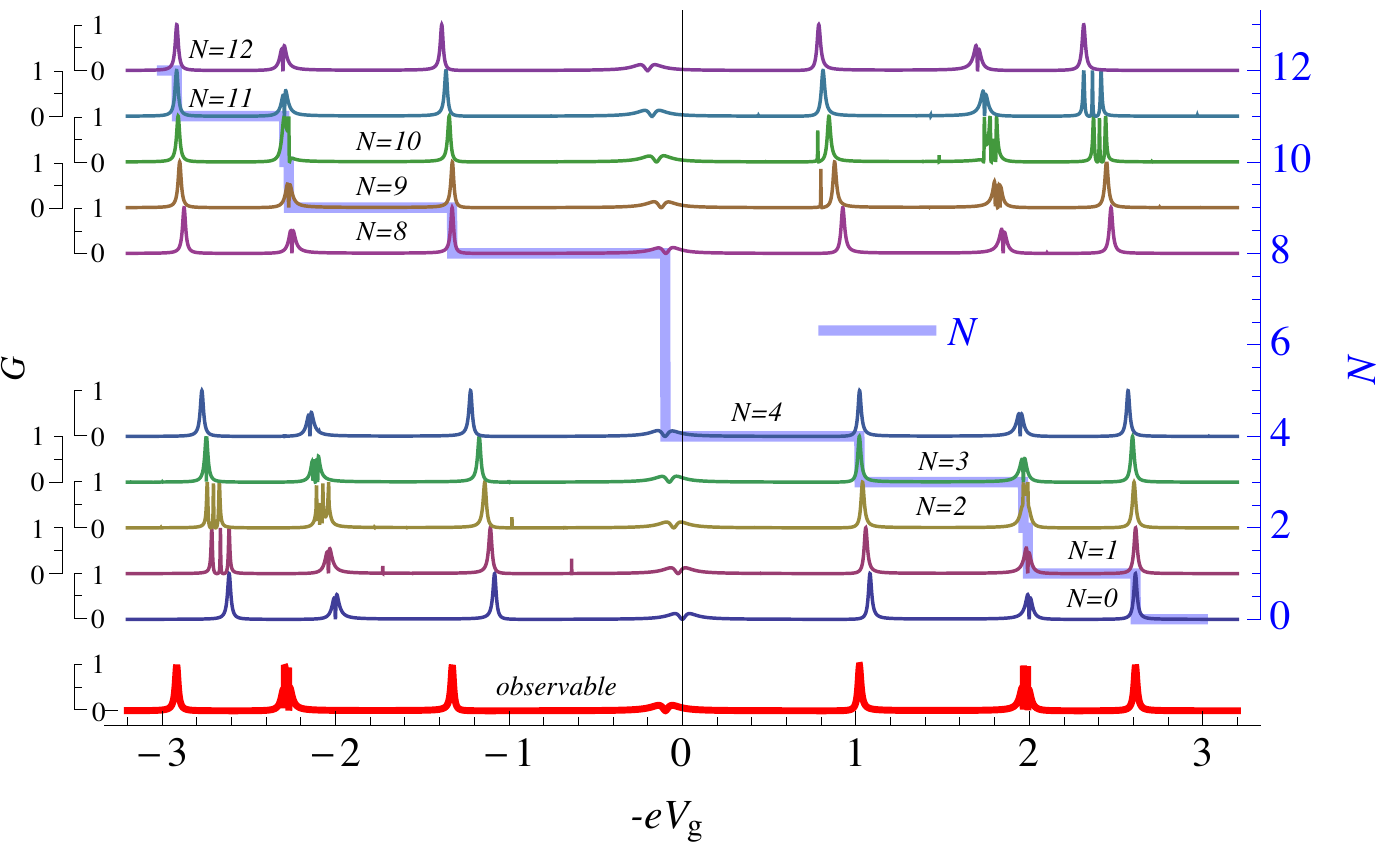}}\\
    \subfloat [$V/t=1$] 
    {\label{fig:GexpresultsV1_B1C1_fluxPiover2}\includegraphics[width=.45 \textwidth]{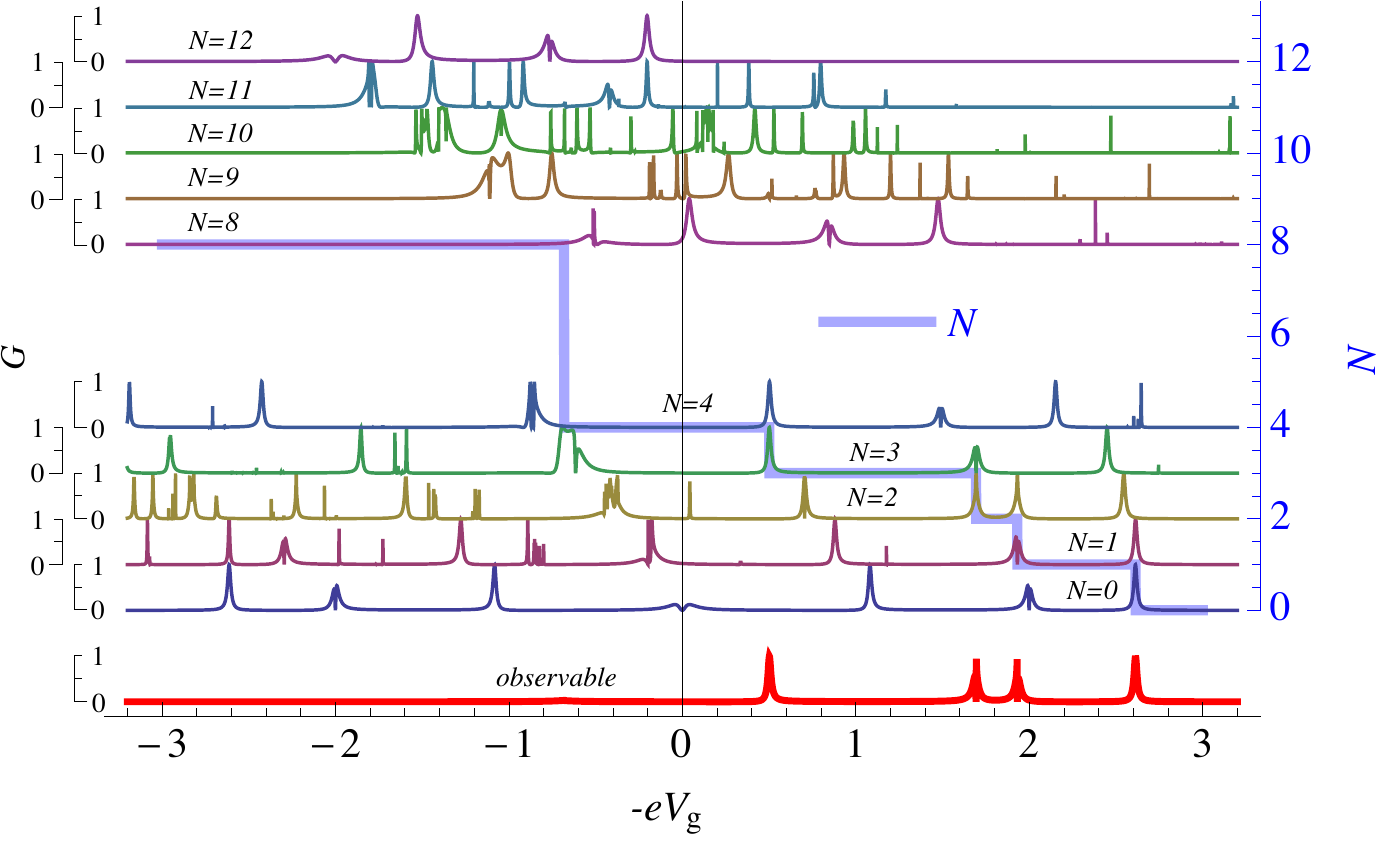}}\\
    \subfloat [$V/t=100$] 
    {\label{fig:GexpresultsV100_B1C1_fluxPiover2}\includegraphics[width=.45 \textwidth]{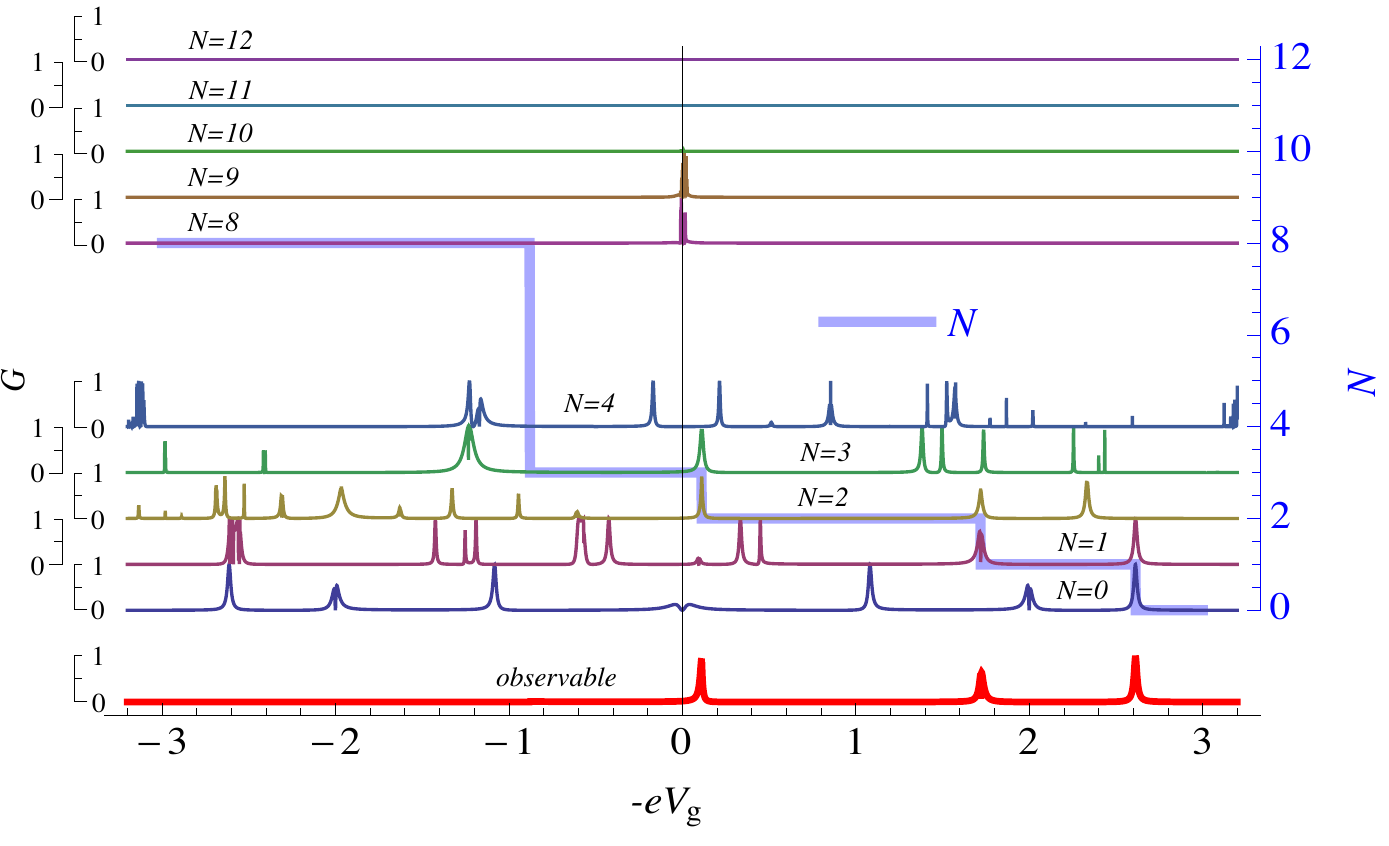}}
    \caption{Conductance profiles in the case of a flux per plaquette $\phi=\pi/2$ (created by an uniform magnetic  field). All other parameters are the same as in Fig.~\ref{fig:Gzeroflux}. An additional peak corresponding to the transition between 3 and 4 particles is present for small and intermediate $V/t$, but absent for large $V/t$.
   }
    \label{fig:Gwithflux}
\end{figure}

\begin{figure}[t]
    \subfloat [AB$_2$ ring] 
     {\label{fig:GexpresultsAAV100}\includegraphics[width=.45 \textwidth]{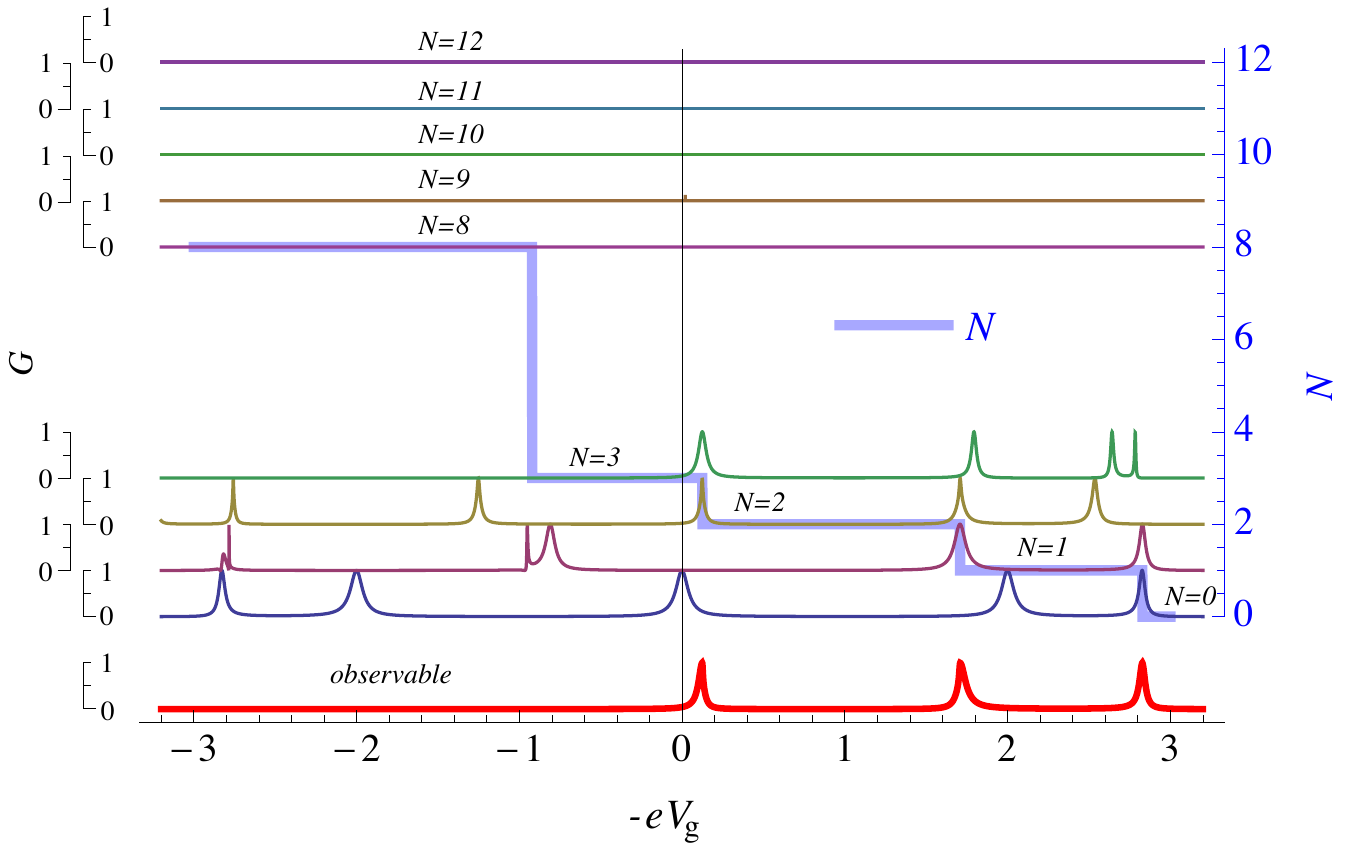}}\\
    \subfloat [linear ring] 
    {\label{fig:GexpresultsringV100}\includegraphics[width=.45 \textwidth]{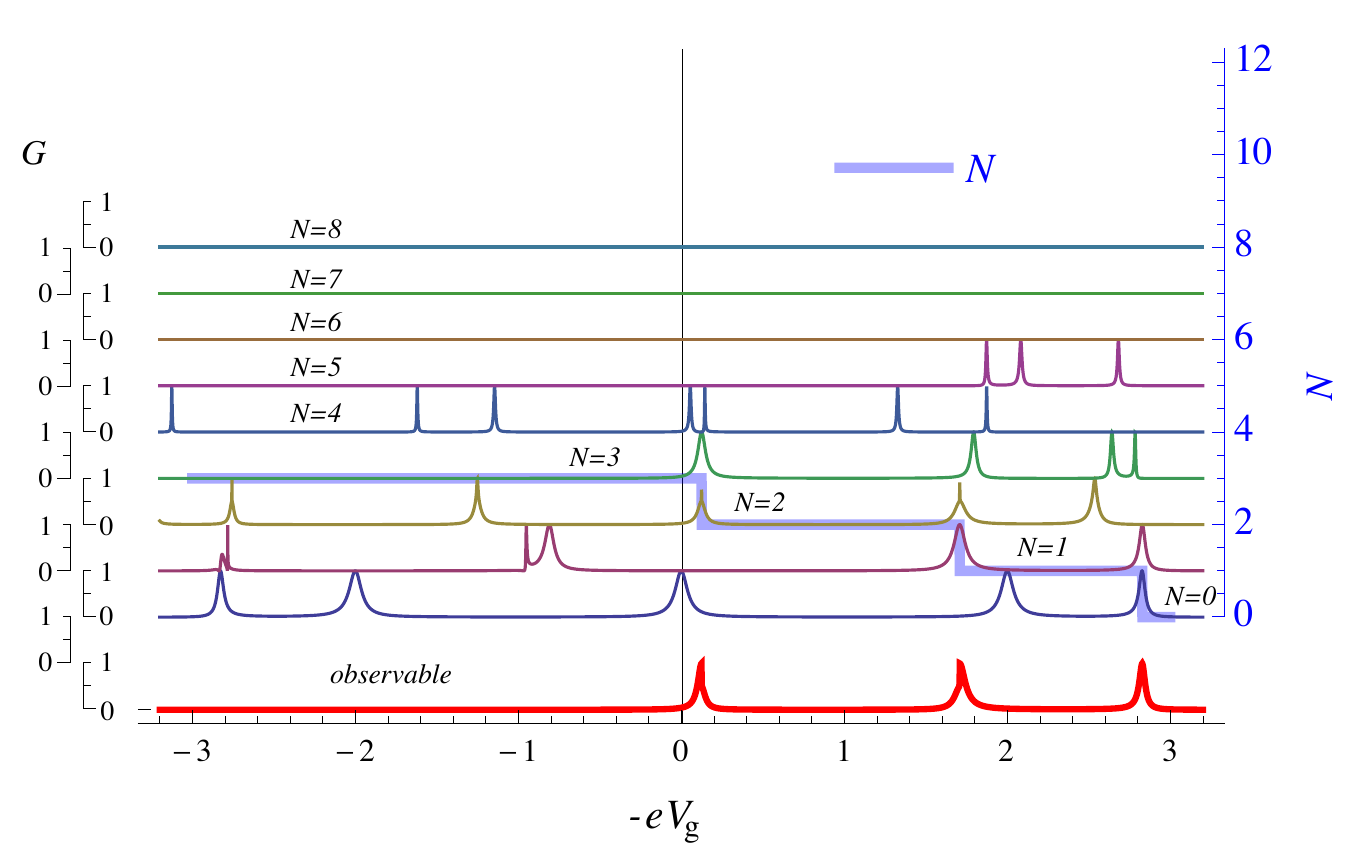}}
    \caption{Conductance  as a function of the gate voltage for: (a) an AB$_2$ ring with $N_c = 4, V = 100, \omega = 0$ and $t_L = t_R = 0.3 t$ with contacts  at opposite sites A; (b) the equivalent linear ring, with the same parameters but ring hoppings renormalized as $t\rightarrow \sqrt{2} t$ and leads contacts  at opposite sites. 
%    The small difference in the two profiles is due to the finiteness of the interaction.
    }
    \label{fig:infiniteVAB2Ring}
\end{figure}

As expected, the conductance profiles obtained using the method of section II reflect closely these phase diagrams, with some peculiarities that we describe below. 
Fig.~\ref{fig:Gzeroflux} shows conductance peaks for a fluxless AB$_2$ chain with conducting leads contacting  sites B and C of the same unit cell, as a function of  the gate potential. In Fig.~\ref{fig:Gzeroflux} as well as in Figs.~\ref{fig:Gwithflux} and \ref{fig:infiniteVAB2Ring}, we show the conductance profiles with fixed number $N$ of particles in the cluster (top curves) as well as the observable conductance profile (bottom red curve) which takes into account the transitions in cluster particle number given by the thick blue curve. The conductance profiles with fixed $N$ are shifted in order to show clearly the observable regions in each of them. Whenever the ground state is degenerate we use the averaging procedure described in section \ref{sec:CTIC}. The chemical potential of the leads is assumed to be 0.

For $V/t=0.1$ (Fig.~\ref{fig:GexpresultsV01}), the conductance profiles are very similar independently of the particle number in the cluster. This reflects the fact that  without interaction, the conductance profiles are the same independently of the number of particles in the cluster and the effect of the small interaction is to shift slightly the peaks and partially lift the degeneracy of the cluster Hamiltonian, therefore splitting some peaks. We observe that the zero frequency dipped peak (discussed in detail in Ref.~\cite{lopes_conductance_2014}) survives for small $V/t$, but slightly shifted in frequency.
For $V/t=1$ (Fig.~\ref{fig:GexpresultsV1}), this dipped peak is absent, but a trace of this peak is still observed at point A of the bottom profile of Fig.~\ref{fig:GexpresultsV1}. This trace  appears to be a very asymmetric peak since it is the union of two tails due to the particle number jump.
In the case of  Fig.~\ref{fig:GexpresultsV100}, we show the conductances profiles for  $V/t=100$ and we see that a conductance peak is not necessarily associated with the energy of the particle number jump. This just means that the jump occurs before the ground state with 3 particles becomes degenerate with the ground state with 4 particles.

In Fig.~\ref{fig:Gwithflux}, we show the conductance profiles in the case of a flux per plaquette $\phi=\pi/2$ (created by an uniform magnetic  field). All other parameters are the same as in Fig.~\ref{fig:Gzeroflux}.
These conductance profiles reflect the existence of the $N=4$ region in the phase diagram of Fig. \ref{fig:densityPlotLegendwithflux} by showing an additional conductance peak when $V/t$ is less than 5.

A peculiar feature of the conductance profiles with fixed $N$ shown in Figs.~\ref{fig:Gzeroflux} and \ref{fig:Gwithflux} is presence of peaks of height 0.5. This is a consequence of the degeneracy of the ground state with $N$ particles and of the existence of a conductance peak  for only one of the degenerate ground states. Since we average the conductance over the possible ground states, this leads to a  lower height conductance  peak associated with transition $N \rightarrow N+1$. Note that this average does not occur when the transition occurs in the direction $N+1 \rightarrow N$ and this leads to   a non-Lorentzian peak as observed in the bottom curve of Fig.~\ref{fig:GexpresultsV100} (peak B). One can also observe in the $N=2$ conductance profile of Fig.~\ref{fig:GexpresultsV100}, a  peculiar non-Lorentzian peak  with a sharp maximum (peak C). As explained in section II, this reflects the averaging of two peaks with different widths  due to a degenerate ground state.

%%%%%%%%%%%%%%%%%%%%% 
%      figure       % 
%%%%%%%%%%%%%%%%%%%%% 
%%%%%%%%%%%%%%%%%%%%% 
%      end          % 
%%%%%%%%%%%%%%%%%%%%%
The strong coupling limit solution given in \cite{lopes_interacting_2011} also indicates that for small particle number  the AB$_2$ ring  should behave as a renormalized ring with hopping $\sqrt{2}t$ and reduced length equal to $2N_c/3$. Fig.~\ref{fig:infiniteVAB2Ring} shows a comparison of the conductance peaks for an incident particle with $k = \pi/2$ as a function of the gate voltage for an AB$_2$ ring and for the corresponding renormalized linear ring with leads at opposite A sites for an interaction $V = 100 t$. As expected, the conductance profiles of the two are almost identical.
However the linear ring does not have jumps in particle number as the gate potential is varied.
Note that the jump from 3 to 8 particles in the AB$_2$ ring can be considered similar to the transition from 3 to 4 particles in the linear ring (every other site occupied in the ring and every pair of B and C sites occupied in the AB$_2$ ring), but the latter leads to a conductance peak while the former may not, as explained above. Also, the transition from 3 to 8  in the AB$_2$ ring occurs much earlier.

%%%%%%%%%%%%%%%%%%%%% 
%      figure       % 
%%%%%%%%%%%%%%%%%%%%% 
%%%%%%%%%%%%%%%%%%%%% 
%      end          % 
%%%%%%%%%%%%%%%%%%%%%

%%%%%%%%%%%%%%%%%%%%%%%%%%%%%%%%%%%%%%%%%%%%%%%%%%%%%%%%%%%%%%%%%%%%%%%%%%
%                                  Conclusion                            %
%%%%%%%%%%%%%%%%%%%%%%%%%%%%%%%%%%%%%%%%%%%%%%%%%%%%%%%%%%%%%%%%%%%%%%%%%%
\section{Conclusion}

In this paper, a new method for the determination of the two-terminal differential  conductance through an interacting cluster has been presented and applied to the case of  the conductance of spinless fermions through an AB$_2$ ring considering nearest neighbors interactions.
This method is exact in two limits: (i) vanishing interactions; (ii) vanishing coupling between leads and cluster, and its 
main approximation is the assumption that  
particles in the leads remain independent and therefore the system can be reduced to a simpler problem of one incoming particle which interacts with $N$ particles in the cluster.
This simpler problem can studied in a truncated Hilbert space and in this space, it  can mapped exactly 
into the problem of the transmission of an incoming particle through a non-interacting cluster of $M$ independent sites (where $M$ is the number of cluster states with one  particle  more or less than the ground state of the cluster).

The main results obtained in our study of the conductance features due to interactions in the AB$_2$ ring are: (i) the non-interacting conductance profiles persist for small values of the interaction, with a small shift of the peaks and a small splitting of the peaks (when degeneracy is present in the non-interacting limit); (ii)  some conductance peaks are not present due to a particle number jump (observed for any value of the interaction) that occurs as the gate potential is varied and that can be associated with the flat band of the AB$_2$ ring when interactions are absent.

%%%%%%%%%%%%%%%%%%%%%%%%%%%%%%%%%%%%%%%%%%%%%%%%%%%%%%%%%%%%%%%%%%%%%%%%%%
%                        Acknowledgements                                 %
%%%%%%%%%%%%%%%%%%%%%%%%%%%%%%%%%%%%%%%%%%%%%%%%%%%%%%%%%%%%%%%%%%%%%%%%%%

\subsection*{Acknowledgements}
A. A. L. acknowledges the financial support of the Portuguese Science
and Technology Foundation (FCT), cofinanced by FSE/POPH, under grant SFRH/BD/68867/2010 and of the Excellence Initiative of the German Federal and State Governments (grant ZUK 43).
R. G. D.  acknowledges the financial support from the Portuguese Science
and Technology Foundation (FCT) through the program PEst-C/CTM/LA0025/2013.

\appendix

%%%%%%%%%%%%%%%%%%%%%%%%%%%%%%%%%%%%%%%%%%%%%%%%%%%%%%%%%%%%%%%%%%%%%%%%%%
%                 Derivation of a formula for the conductance            %
%%%%%%%%%%%%%%%%%%%%%%%%%%%%%%%%%%%%%%%%%%%%%%%%%%%%%%%%%%%%%%%%%%%%%%%%%%
\section{Derivation of a formula for the conductance}
\label{app:DFC}

In this section we provide a derivation of the transmittance of spinless fermions through an interacting  cluster connected to one-dimensional tight-binding semi-infinite leads. Our results for the transmitance agree with those of \cite{enss_impurity_2005} for the non-interacting case, while we have checked numerically that they are consistent with those of \cite{rincon_features_2009} in the interacting case. 

Let us begin by fixing our notation. We let $c_i^\dagger\ket{\emptyset_{L,R}}$ denote the one-particle Wannier state of the leads corresponding to a particle at site $i$ (with  $\ket{\emptyset_{L,R}}$ being the vacuum state of the left or right lead) and $\ket{\text{GS}(N)}$ denote the ground state of the cluster with $N$ particles. Whenever we consider the tensorial product of three states, $\ket{\cdot} \otimes \ket{\cdot} \otimes \ket{\cdot}$, the first state is a left lead state, the second one is right lead state and the last one a cluster state.

We consider two semi-infinite leads connected to a cluster as depicted in Fig.~\ref{fig:DiamondStar4SidedLeads}. In particular we take the Hamiltonian for our system to be
\begin{equation}
    H = H_0 + V_{LR},
\end{equation}
where $H_0$ is the Hamiltonian of the isolated cluster plus leads,
\begin{equation}
    \begin{split}
        H_0 &= -t\sum_{j=-\infty}^0 \left( c_{j-1}^\dagger c_j + \text{H.c} \right) \\
            &= -t\sum_{j=N+1}^\infty \left( c_j^\dagger c_j + \text{H.c} \right) \\
            &+ H_c,
    \end{split}
\end{equation}
where $H_c$ is the Hamiltonian of the cluster
and $V_{LR}$ is the hybridization between the cluster and the leads, taken as a perturbation,
\begin{equation}
    V_{LR} = -t_L c^\dagger_0 c_L - t_R c_R^\dagger c_{N+1} + \text{H.c.} .
\end{equation}

As explained in section II, 
 we  study the  transmission of a single incoming particle through the interacting cluster following a Landauer-like procedure assuming that the particles in the leads remain independent despite interacting with particles in the cluster.
Furthermore, the number of particles in the cluster is fixed by the chemical potential in the leads and the  electrons in the lower energy states of the cluster do no tunnel to the leads due to the Pauli exclusion principle.
Let us assume  that the chemical potential is adjusted so that  the ground state of the cluster has exactly $N$ electrons.
Considering this simpler problem of a single particle in the leads and $N$ particles in the cluster, one is lead to the conclusion that when the particle in the leads is far from the cluster, the ground state of the cluster is  the same as that of the decoupled system. Also, the incoming particle will be  in a plane wave state (since these are eigenstates of the leads Hamiltonian and we take the hybridization to be small).

So, when the incoming particle is far in the left lead ($j \ll 0$), the particle+cluster eigenfunction for a certain energy $\omega_k +E_{GS(N)}$  will be  a combination of a incident plane wave  with a reflected  plane wave due to the cluster
\begin{equation}
    \left( \e^{\I k j} + \psi_r \e^{-\I k j} \right) c_j^\dagger \ket{\emptyset_L} \otimes \ket{\emptyset_R} \otimes \ket{\text{GS(N)}} ,
\end{equation}
while when the particle is far in the right lead  ($j \gg N+1$) one has a transmitted component
\begin{equation}
    \ket{\emptyset_L} \otimes \psi_t \e^{\I k j}c_j^\dagger \ket{\emptyset_R} \otimes \ket{\text{GS(N)}}.
\end{equation}
The previous expressions of the eigenstate with energy $\omega_k +E_{GS(N)}$ can be extended to 
$j \leq 0$ and  $j\geq N+1$, respectively, applying the induction method starting  from these far away components [as for the non-interacting case, cf. \cite{antonio_transport_2013} eq. (11)] using,
\begin{equation}
    \omega_k \psi_j =  -  \psi_{j+1} - \psi_{j-1},
    \label{tbEq}
\end{equation}
That is, solving successively the previous matrix equation, we can  get closer to the cluster, starting from the $j \rightarrow -\infty$  and $j \rightarrow \infty$ cases.

There are two possibilities for the transmission of the incoming particle through the cluster, one  involving an intermediate  state with $N+1$ particles in the cluster and zero particles in the leads and another  involving an intermediate  state $N-1$ particles in the cluster and one particle at site 0 and one particle at site $N_s+1$. Taking into account the above discussion, we restrick our analysis to the following subspace of states
\begin{equation}
    \begin{split}
        c_j^\dagger\ket{\emptyset_L} \otimes \ket{\emptyset_R} \otimes  &\ket{\text{GS}(N)} , \quad j \leq 0,  \\
        \ket{\emptyset_L} \otimes c_j^\dagger \ket{\emptyset_R} \otimes &\ket{\text{GS}(N)} , \quad j \geq N+1, \\
        \ket{\emptyset_L} \otimes \ket{\emptyset_R} \otimes &\ket{N+1_{(n)}},                                        \\
        c_0^\dagger \ket{\emptyset_L} \otimes c_{N+1}^\dagger \ket{\emptyset_R}\otimes &\ket{ N-1_{(m)}}, 
    \end{split}
    \label{subspace}
\end{equation}
where $\ket{N+1_{(n)}}$  represents all  cluster states with $N+1$  particles and $\ket{ N-1_{(m)}}$ represents  cluster states with $N-1$  particles. The reduced Hilbert space implies that when the particle in the leads is far from the cluster, the ground state of the cluster is  the same as that of the decoupled system, which is a valid approximation given that we assumed that the hybridization between the leads and the cluster is small.
\begin{widetext}

Let us now write the full form of the eigenstate with energy $\omega_k + E_{GS, N}$, not forgetting that we are working in a restricted subspace (we are in fact, simply expanding this state in terms of the basis for our subspace, using the knowledge we have of how the eigenstate on the leads must look like)
\begin{equation}
    \begin{split}
    \ket{\psi_k} &= \sum_{j\leq 0} \left( e^{\I k j} + \psi_r \e^{-\I j k} \right) c_j^\dagger \ket{\emptyset_L} \otimes \ket{\emptyset_R} \otimes \ket{\text{GS(N)}} 
                 + \sum_{j\geq N+1} \psi_t \e^{\I k j} \ket{\emptyset_L} \otimes c_j^\dagger \ket{\emptyset_R} \otimes \ket{\text{GS(N)}} \\
                 &+ \sum_{n} \alpha_n \ket{\emptyset_L} \otimes \ket{\emptyset_R}  \otimes  \ket{N+1_{(n)}}  
                 + \sum_{m}  c_0^\dagger \ket{\emptyset_L} \otimes c_{N+1}^\dagger \ket{\emptyset_R} \otimes \beta_m \ket{N-1_{(m)}},  
   \end{split}
\end{equation}
where $\ket{N+1_{(n)}}$ denotes the $n$th eigenvector of the cluster Hamiltonian when it has $N+1$ particles and identically for $\ket{N-1_{(m)}}$.
Let us now look at the Hamiltonian matrix equations which involve $V_{LR}$.
For $c_0^\dagger \ket{\emptyset_L} \otimes \ket{\text{GS(N)}} \otimes \ket{\emptyset_R}$ we have
\begin{equation}
    \begin{split}
       \left( \omega_k + E_\text{GS(N)} \right) \psi_0 c_0^\dagger \ket{\emptyset_L} \otimes \ket{\emptyset_R} \otimes \ket{\text{GS(N)}}  
       = &  -\psi_{-1} c_0^\dagger \ket{\emptyset_L} \otimes \ket{\emptyset_R} \otimes \ket{\text{GS(N)}} \\
        &  - t_L c_0^\dagger \ket{\emptyset_L} \otimes \ket{\emptyset_R} \otimes \sum_i \left(  \alpha_i c_L\ket{N+1_{(i)}} \right)  \\
        &-t_R c_0^\dagger \ket{\emptyset_L} \otimes \ket{\emptyset_R} \otimes \sum_j \left(  \beta_j c_R^\dagger \ket{N-1_{(j)}} \right)  \\
        &+ E_\text{GS(N)} \psi_0 c_0^\dagger \ket{\emptyset_L} \otimes \ket{\emptyset_R} \otimes \ket{\text{GS(N)}}.
    \end{split}
\end{equation}
% For simplicity we now drop the $\ket{\emptyset}$ factor which is always present.
We emphasize that we work in a restricted subspace where only one state with $N$ particles in the cluster is available, the ground state, so a projector is implicit in the previous equation. The degenerate ground state case can be treated by using the Gibbs state as described in section \ref{sec:CTIC}. Given this, and through some straightforward calculations we can arrive at
\begin{equation}
    \begin{split}
        \omega_k \psi_0 + \psi_{-1} = & -t_L \sum_n \alpha_n \bra{\text{GS}(N)} c_L \ket{N+1_{(n)}} 
        -t_R \sum_m \beta_m  \bra{\text{GS}(N)} c_R^\dagger \ket{N-1_{(m)}} \\
       = & - \sum_n \left(
        t_{\alpha_n(N)}^{L} 
        \right)^* \alpha_n  
        -\sum_m \left(
        t^{R}_{\beta_m(N)} \right)^* \beta_m.  
    \end{split}
                                    \label{eq:1}
\end{equation}
Now, for $\ket{\emptyset_L} \otimes \alpha_n \ket{N+1_{(n)}} \otimes \ket{\emptyset_R}$ after some calculations we get

\begin{equation}
    \begin{split}
        \left( \omega_k + E_\text{GS}(N) - E_{\alpha_n}\right) \alpha_n 
        = & - t_L \psi_0 \bra{N+1_{(n)}} c_L^\dagger \ket{\text{GS}(N)} 
         - t_R \psi_{N+1} \bra{N+1_{(n)}} c_R^\dagger \ket{\text{GS}(N)}\\
        = & - \left( t_{\alpha_n(N)}^{L}  \right) \psi_0   - \left(
       t^{R}_{\alpha_n(N)} \right)^* \psi_{N+1}.
    \end{split}
    \label{eq:2}
\end{equation}
For $c_0^\dagger \ket{\emptyset_L} \otimes \beta_m \ket{N-1_{(m)}} \otimes c^\dagger_{N+1} \ket{\emptyset}$ we get

\begin{equation}
    \begin{split}
        \left( \omega_k + E_\text{GS}(N) - E_{\beta_m} \right) \beta_m =& + t_L \psi_{N+1} \bra{N-1_{(m)}} c_L \ket{\text{GS}(N)} 
        - t_R \psi_{0} \bra{N-1_{(m)}} c_R  \ket{\text{GS}(N)} \\
        = & - \left( t_{\beta_m(N)}^{L}  \right)^* \psi_{N+1}   - \left(
       t^{R}_{\beta_m(N)} \right) \psi_{0}.
     \end{split}
        \label{eq:3}
\end{equation}
Finally for $\ket{\emptyset_L} \otimes \ket{\text{GS}(N)} \otimes c_{N+1}^\dagger\ket{\emptyset}$ we have

\begin{equation}
    \begin{split}
        \omega_k \psi_{N+1} + \psi_{N+2} =& -t_R \sum_n \alpha_n \bra{\text{GS}(N)} c_R \ket{N+1_{(n)}} 
        + t_L \sum_m \beta_m \bra{\text{GS}(N)} c_L^\dagger \ket{N-1_{(m)}}\\
         = & - \sum_n \left( t_{\alpha_n(N)}^{R}  \right) \alpha_n   - \sum_m \left(
         t^{L}_{\beta_m(N)} \right) \beta_m.                                   
    \end{split}
                                            \label{eq:4}
\end{equation}
\end{widetext}
Note we have introduced the parameters
\begin{eqnarray}
t_{\alpha_n(N)}^{L} & =&  t_L  \bra{N+1_{(n)}} c_L^\dagger \ket{\text{GS}(N)},  \\
t_{\alpha_n(N)}^{R} & =& t_R \bra{\text{GS}(N)} c_R \ket{N+1_{(n)}} , \\
t^{R}_{\beta_m(N)} & =& t_R \bra{N-1_{(m)}} c_R  \ket{\text{GS}(N)},  \\
t^{L}_{\beta_m(N)} &= & -t_L  \bra{\text{GS}(N)} c_L^\dagger \ket{N-1_{(m)}}.
\end{eqnarray}
Defining 
\begin{eqnarray}
\varepsilon_{\alpha_n(N)} & =&   E_{\alpha_n} - E_\text{GS}(N), \\
\varepsilon_{\beta_m(N)} & =& E_{\beta_m} -E_\text{GS}(N),
\end{eqnarray}
one concludes that  Eqs.~\ref{eq:1}, \ref{eq:2}, \ref{eq:3} and \ref{eq:4} correspond to the Hamiltonian matrix equations  of the effective system shown in Fig.~\ref{fig:effectcluster}.

The solution of this set of $M+2$ equations (that is, Eqs.~\ref{eq:1}, \ref{eq:2}, \ref{eq:3} and \ref{eq:4}) allows us to determine $\psi_t$ and $\psi_r$. Note that $\psi_0$, $\psi_{-1}$, $\psi_{N+1}$ and  $\psi_{N+2}$ are given by expressions $\psi_j=\left( \e^{\I k j} + \psi_r \e^{-\I k j} \right)$ for $j \leq 0$ and  $\psi_j=\psi_t \e^{\I k j}$ for $j \geq N+1$ and 
are  functions of $\psi_r$ and $\psi_t$, therefore we have $M+2$ variables. 
The transmission probability  is then given by the square of the absolute value of the ratio between the   amplitude of the  outgoing wave $\psi_t$ and the amplitude of the incident wave (which we have assumed to be 1). 

%%%%%%%%%%%%%%%%%%%%%%%%%%%%%%%%%%%%%%%%%%%%%%%%%%%%%%%%%%%%%%%%%%%%%%%%%%
%                       AB$_2$ chain                                     %
%%%%%%%%%%%%%%%%%%%%%%%%%%%%%%%%%%%%%%%%%%%%%%%%%%%%%%%%%%%%%%%%%%%%%%%%%%
\section{Exact spectrum  of the AB$_2$ chain}

\label{app:ESAB2C}

In this appendix, we   recall results on the exact diagonalization of the spinless AB$_2$ chain taking into account nearest-neighbor Coulomb interactions in the limiting cases of infinite or zero nearest-neighbor Coulomb repulsion for any filling.

For the non interacting case, the one-particle eigenvalues for an arbitrary value of flux $\phi$ are  given by
\begin{equation}
        \begin{split}
            \epsilon_\text{flat} &= 0, \\
            \epsilon_{\pm} &= \pm 2t \sqrt{1+ \cos(\phi/2)\cos(\phi^\prime/N_c + k)}.
        \end{split}
\end{equation}

In the strong coupling limit, $V/t \rightarrow \infty$, the solution to the AB$_2$ chain can be derived from the solution to the t-V chain \cite{dias_exact_2000}.
The solution of the t-V chain relies in 
considering pairs of consecutive sites, nearest neighbors of each other. The possible pairs, which we call links, are $(h, p)$, $(p, h)$, $(h, h)$ and  $(p, p)$, where $p$ stands for an occupied site and $h$ for an empty one. In the strong coupling limit, there is  a conservation of the number of these links, so that the tight-binding term only exchanges the position of these links \cite{dias_exact_2000}. 
Considering the subspace where there are no pairs of nearest neighbor occupied and interpreting the $(h,p)$ links as non-interacting particles hopping in a chain where the empty sites are the $(h, h)$ links, the solution is attained.
Since the total number
of $(hh)$ and $(hp)$ links is $\tilde{L}=L-N$, the effective chain is reduced in relation to the real t-V chain. 
The fact that the tight-binding particles occupy two sites of the real t-V chain leads to a twisted boundary condition which is dependent on the momenta of the tight-binding particles and the  eigenvalues are given by \cite{dias_exact_2000}
\begin{equation}
        E(\{\tilde{k}\},P)=-2 \sqrt{2} t \sum_{i=1}^{N}
        \cos \left(\tilde{k}_i-\dfrac{P}{\tilde{L}}  -{\phi^\prime \over L} \right),
        \label{eq:sevE}
\end{equation}
with $\tilde{k}=\tilde{n} \cdot 2\pi /\tilde{L}$,
and $P=n \cdot 2\pi /L$, with $\tilde{n}=0, \dots, \tilde{L}-1$
and $n=0, \dots, L-1$.
The set of  pseudo-momenta
$\{\tilde{k}\}$ and $P$  must satisfy the  following condition  
$
        P L/\tilde{L}=\sum_{i=1}^{N} \tilde{k}
        \quad (\text{mod} \,\, 2\pi).
$
A similar reasoning gives the energy of the states with $(p, p)$ links. 
The mapping of this solution of the t-V chain (with even number of sites) into the AB$_2$ chain without localized particles  is direct,
with odd sites corresponding to A sites and even sites to sites B and C (which will be unoccupied or in a bonding configuration). 
This entails a reduction of the effective number of sites, which can be seen in the conductance of the AB$_2$ chain.

In this strong-coupling limit, two kinds of localized states may occur: one-particle localized states due to geometry and two-particle localized states due to interaction and geometry. These  localized fermions create open boundary regions for itinerant carriers.
This implies a curious dependence of the ground state energy as function of filling.In the case of an infinite AB$_2$ chain, at filling $\rho=2/9$ and in order to avoid the existence of itinerant fermions  with positive kinetic energy, phase separation occurs between a high-density phase ($\rho=2/3$) and a low-density phase ($\rho=2/9$) leading to a  metal-insulator transition. The ground-state energy is linear on filling above $2/9$.

\bibliography{prb}

\end{document}